# HEALTHCARE IT: IS YOUR INFORMATION AT RISK?

Kimmarie Donahue[1] and Syed (Shawon) M. Rahman, PhD[2]

[1]Information Assurance Project Lead, San Antonio, TX, USA
`KDonahue@CapellaUniversity.edu`
[2]Assistant Professor, University of Hawaii-Hilo, Hilo, USA
and Adjunct Faculty, Capella University, Minneapolis, USA
`SRahman@Hawaii.edu`

*ABSTRACT*

*Healthcare Information Technology (IT) has made great advances over the past few years and while these advances have enable healthcare professionals to provide higher quality healthcare to a larger number of individuals it also provides the criminal element more opportunities to access sensitive information, such as patient protected health information (PHI) and Personal identification Information (PII). Having an Information Assurance (IA) programallows for the protection of information and information systems andensures the organization is in compliance with all requires regulations, laws and directive is essential. While most organizations have such a policy in place, often it is inadequate to ensure the proper protection to prevent security breaches. The increase of data breaches in the last few years demonstrates the importance of an effective IA program. To ensure an effective IA policy, the policy must manage the operational risk, including identifying risks, assessment and mitigation of identified risks and ongoing monitoring to ensure compliance.*

## 1. INTRODUCTION

Advances in today's Healthcare Information Technology have allowed healthcare professionals to become highly connected to the information highway which provides them greater access to patients and their healthcare information. In today's society it is becoming more and more common to see healthcare professionals to utilizing mobile devices, such as laptops, to allow them to be better connected, according to a recent survey conducted, over 80 percent of Healthcare IT professionals allow IPads on the enterprise network and 65 percent provide support for iPhones and iPod Touch devices [1].

While these advances have provided great benefits for healthcare professionals and their patients, they also pose a real danger not only to the patients protected heath information (PHI) but also the organizations that are affected by data breaches. Healthcare data breaches are up by ninety seven percent in 2011; this is usually due to malicious attacks such as, theft of laptops, carelessness of an insider threat or hacking[2]. A study from the Ponemon Institute estimated the cost of data breaches has increased for the fifth year in a row to $7.2 million dollars and costs organizations an average of $214 per compromised record[3]. Medical ID theft is becoming big business, the World Privacy Forum found that a social security number has a street value of one dollar and a stolen medical identity goes for fifty dollars [4].

Once the dangers have been identified the next step is to ensure that senior management understands the risk if nothing is done to manage the threats and vulnerabilities. Typically security is an afterthought and organizations are reluctant to budget accordingly, until a security breach occurs.

Information Assurance refers to the measures that organizations take to protect and defend not only information but also information systems by ensuring their availability, integrity, authentication, confidentiality and non-repudiation [5].

Most individuals expect Healthcare professionals and IT professionalsto uphold a higher ethical and legal standard due to their access to sensitive information required in their daily responsibilities and as such should always practice due care and due diligence. It is important that management stay in touch with their personnel, there are personnel that may usually maintain a high ethical standard, but are often easily be influenced if the right opportunity presented itself, such as pending layoff or financial difficulties. Cyber-Ark, a security firm, conducted a survey called "The Global Recession and its Effects on Work Ethics" which revealed that of the individuals interviewed, 56% of workers worried about the loss of their jobs and over half admitted they downloaded sensitive data in order to try to use it at their next position [6].

If an organization fails to practice due care or due diligence,they will be accountable, financially and/or criminally, especially if the threat could have been avoided. A data breach in California resulted in the California Department of Health fining several California hospitals $675,000 for repeatedly failing to adequately secure patient data,and in Louisville a university hospital physician hospital inadvertently exposed the personal information of over 700 patients receiving kidney dialysis treatment when he set up the patient database on an unsecure website [7]. In 2010 CPA Tim W. Kasley was disbarred for failure to exercise due diligence while he was preparing tax returns for a corporation, he failed to determine the right information he received for the tax return [8]. This emphasise the need for all individuals in an organization to exercise due care and due diligence.

- Performing Due care:
    - Taking responsibility when identifying a potential threat or risk and having the responsibility to know or find out what actions will correct or eliminate the threat or risk.

- Performing Due diligence:
    - Taking the responsibility to put controls in place and properly monitor to mitigate or eliminate the threat or risk, and perform risk analysis as required.

Establishing a comprehensive information assurance program will ensure that all individuals understand the importance to ensuring the security of not only the enterprise but also the sensitive information on accessed on the enterprise. Organizations must ensure that that not only is a Information Assurance program is in place but also that it is adequate enough to address the increased threats to the confidentiality, integrity, and availability of sensitive information, such as patient health information, and stays in compliance with all financial, legal and health care compliance regulations. To ensure the success of any IA program it is essential that senior management fully endorse the program, because if senior management does not support it then no one else will support it.

An important part of an IA program is building the policies that will help provide for an Defense in Depth approach to IA by providing layers of principles and controls that cover not only the individuals but also the various process and technology the organization uses, including roles and responsibilities, acceptable use, etc.

Risk management is a critical part of any IA program because new threats and vulnerabilities are emerging every day. Risk management helps ensures that thesethreats and vulnerabilities are properly identified, and mitigated to reduce risk. It is impossible to eliminate all risk from an enterprise, but mitigated to a level the organization is willing to accept. A risk assessment will provide for the identification of potential threats and vulnerabilities as well as possible mitigation strategies that will bring the identified threats and vulnerabilities to an acceptable level.

Risk management is a continuous process that requires monitoring and updates to ensure that the proper protection and ensure effectiveness and compliance with laws, regulations and directives.

Kingdom Hospital is a factious hospital that is used for this case study and as a hospital has unique requirements, such as medical devices, wireless devices (tablets, blackberries, etc.), Health Insurance Portability and Accountability Act (HIPAA) and Privacy issues that are not currently being fully met. This increases the threat to the confidentiality, integrity, and availability of Kingdom resources and assets, such as electronic protected health information (PHI).

## 2. INFORMATION ASSURANCE PROGRAM

When developing and implementing an IA program it is essential that senior management fully supports and is committed to the program, because without their support and direction the program will not be successful. It is important to develop the policies with input from all business owners to ensure that the IA policies and procedures will not only protect IT resources, but also align themselves with the organizations business objectives. The policies should also be developed so they are easily understood, because is the policies are not understood by all individuals they will just be ignored. The IA plan must be based upon the mission and business objectives of the organization and support the future direction of the organization in order to be successful [9].

### 2.1. ETHICAL AND LAWS IMPLICATIONS

Everyone has a responsibility when it comes to information awareness. Senior management has the key responsibility to not only support and promote the IA program to the organization but also to ensure that the organization is in compliance with the industry laws and regulations, such as Privacy act, Health Insurance Portability and Accountability Act (HIPAA), etc., because a data breach can be costly for an organization. Potential lawsuits resulting from data breaches could result in big losses, such as the Emory Healthcare in Atlanta, which has a pending class action lawsuit, resulting from a data breach that compromised the personal information of an estimated 315,000 patients, which could cost the organization an estimated $200 million [10].

## 3. ESTABLISH INFORMATION ASSURANCE POLICIES

### 3.1. ACCEPTABLE USE

A key element of any IA policy is an acceptable use policy. This policy will establish what behaviour is appropriate and acceptable by the organization; this includes what the individual is/is not allowed to do, but also covers the consequences for noncompliance with the policy. Recommended practice is to have each individual sign an acceptable use policy agreement; this not only helps to minimize potential legal action but also helps ensure compliance with industry laws and regulations, such as privacy and HIPAA.

### 3.2. TRAINING AND AWARENESS

An essential element of an IA Policy is the training and awareness program and an essential part if an effective IA program is the personnel; they are usually the first line of defense when it comes to protecting information and information systems. Individual actions, intentional or unintentional, greatly contribute to the loss of data breaches. In April 2012 alone three data breaches resulting in almost 1.1 million records being lost, these breaches were caused by the actions of an insider threat, actions while unintentional, still had devastating results[11]. Protecting against this type of threat can be challenging because the individuals have access to the data may not fully understand the impact of their actions. This is why training is essential, to educate on potential threats and vulnerabilities that exist, but also how to recognise them and react appropriately when faced with the threat. IA training should be specialized and focused depending on the audience, such as general users, managers or IT/IA staff (Fig 1).

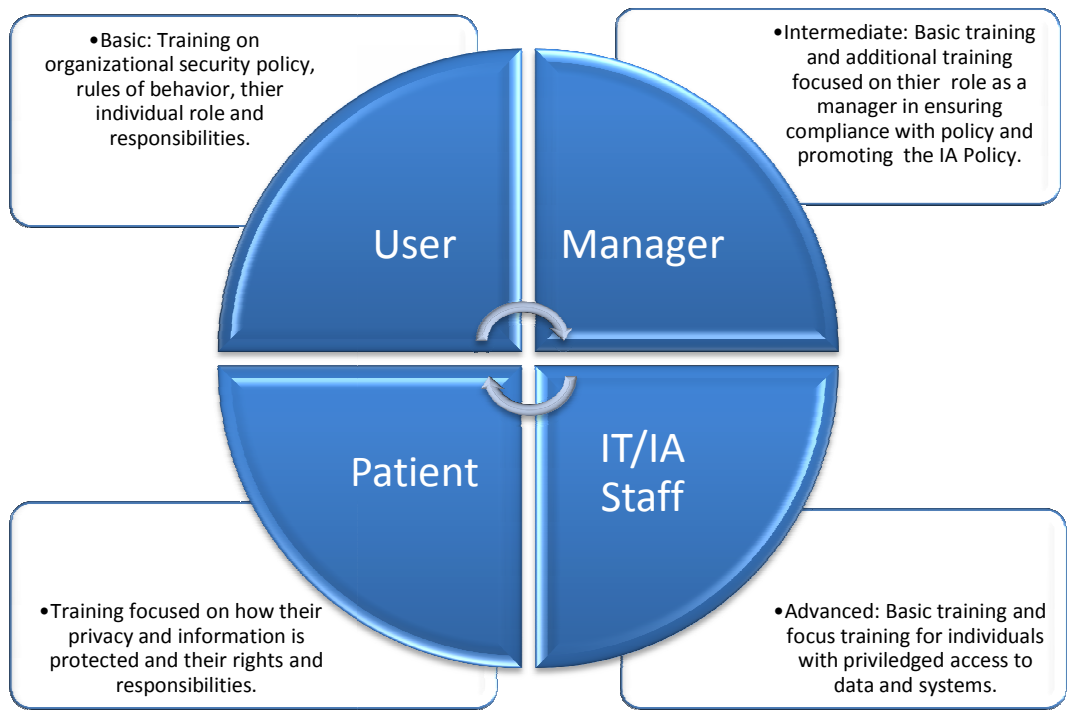

Figure 1: Specialised IA Training Requirements

To ensure training is most effective it should conducted within 30 days of personnel arriving at the organization and annually as a refresher. This initial training provides a basic understanding of various IA concept and principles to ensure the confidentiality, integrity, authentication and availability of the organizations resources and assets. In addition to the basic IA training requirement, it is important to establish more focused IA training based on individuals specific roles within the IA program, such as managers and staff.

IA training, depending on role, should include, but not limited to, the following concepts and principles;

User training:

- Basic understanding of IA concepts,
- Physical security of computer hardware and software,
- Proper protection, handling, storage and access of information,
- Privacy Act, HIPAA and PII protection,
- Recognizing and responding to common threats, vulnerabilities and risks,
- Understanding policy for non compliance to the IA rules and regulations,
- Users should understand their role and responsibility in ensuring the organization's security posture.

Manager training;

- User training,
- Intermediate understanding of IA concepts,
- Intermediate understanding of threats, vulnerabilities and risk,
- Intermediate understanding of governing regulations, laws and directives,
- Managers should understand their role in supporting, promoting and ensuring all users comply with organizations security posture.

IT/IA staff training

- User training,
- Intermediate understanding of IA concepts,
- Intermediate understanding of threats, vulnerabilities and risk,
- Intermediate understanding of governing regulations, laws and directives,
- Staff members with privileged access must understand their role in supporting the organizations security posture but also the added responsibility to act ethically and legally when using their privileged access to IT resources and assets.

Patients Training;

- This is usually in the form of a pamphlet or a handout that informs the patient their role and rights regarding privacy, and how the organization protects their privacy and sensitive information.

To develop and maintain the best possible IA workforce it is essential that individuals, according to the role they are in, train, achieve and maintain appropriate certifications. This allows organizations to ensure that individuals will possess the required knowledge and skills to best perform their individual roles. The following table shows the basic recommended certifications and training for IT workforce members.

Table 1: Recommended IA staff certifications/training

| LEVEL | CERTIFICATION | TRAINING |
|---|---|---|
| BASIC TECHNICIAN | A+, OS | COMPTIA, MICROSOFT, CISCO |
| SENIOR TECHNICIAN | SECURITY+, MCSE | COMPTIA, MICROSOFT |
| NETWORK TECHNICIAN | CCNA | CISCO |
| BASIC IA LEVEL | SECURITY+ | COMPTIA |
| IA MANAGER | CISSP | ISC2 |
| AUDITOR | CISA | ISACA |

## 4. RISK MANAGEMENT

In order to accomplish the goal of protecting an organizations network and information infrastructure from potential compromise or loss, risk management provides a framework for identifying, assessing and mitigating risk down to an acceptable level [12].

### 4.1. RMF

The National Institute of Standards and Technology (NIST) provides a dynamic approach to risk management by allowing for the ability to effectively manage security risks in environments that deal with complex threats and vulnerabilities and rapidly changing missions [13]. This framework allows and organizations to not only assess the risks to their information resources but also to select the best security controls that protects individuals and information systems and also aligns with the organizations business objectives. This risk management framework consists of six steps; categorize, select, implement, assess, authorize and monitor.

#### 4.1.1. CATEGORIZE

An organizations senior management cannot ensure the protection of their information and information systems unless they have a full understanding of what exists in their organization. This is why it is an important start to the risk management process to perform the categorization process; this provides a way to determine the sensitivity and critical nature of the information that resides on their information systems. This allows a decision to be determined on the systems risk level based on how critical the information system is and what the impact to the organization, such as financial, legal, etc, would be to the organization if the system was lost or compromised.

#### 4.1.2. SELECT

Once the categorization process has been completed the determination can be made as to what security controls would be best implemented on the information system to protect the information systems confidentiality, integrity, authentication and availability. The security controls must be able to mitigate the systems risk and not interfere with the organization's mission and ability to function. The implemented security controls must be cost effective; it is not efficient to implement a security control that costs $10,000 for an information system that was determined to be a non-critical system.

### 4.1.3. IMPLEMENT

Once the appropriate security controls have been determined, it is necessary to implement them into the organizations security plan. This will address the documentation, including how the security control will be implemented and with what security configuration settings applied.

### 4.1.4. ASSESS

The assessment of the security controls will to determine if the control is effective. This ensures that the security control was not only implemented properly, meets the security objectives and expectations, and is cost effect for the organization.

### 4.1.5. AUTHORIZE

The authorization to operate is the approval that the systems documentation, risk assessment and overall system is determined to be at an acceptable level for the organization.

### 4.1.6. MONITOR

To ensure the continued effectiveness of implemented security controls continuous monitoring is essential. It is essential to monitorand assess the systems security controls to ensure that any changes in the configuration or updates, this ensures that thesystems security is still effective.

## 5. MONITOR/UPDATE TO ENSURE COMPLIANCE

The best built IA program can be implemented but is it is not constantly assessed and updated then it will become useless and give an organization a false sense of security. New threats and vulnerabilities are not the only concern, new system updates and/or configuration changes occur that could have an impact on security. This is why constant monitoring and updating as required are essential to maintain security and compliance and provide the organization a way to evaluate the effectiveness of the IA program.

There are several ways that assist an organization in this effort, such as a change configuration management board, automated tools to monitor systems on the enterprise and security assessments to help evaluate changes to the systems or the operations environment.

## 6. CASE STUDY: KINGDOM HOSPITAL

Kingdom Hospital is a hospital that realizes the importance of ensuring for the protection of its information resources and assets, including personnel, services and systems. Kingdom hospital has unique requirements, such as medical devices, wireless devices (tablets, blackberries, etc.), Health Insurance Portability and Accountability Act (HIPAA) and Privacy act requirements that not currently fully met. To ensure the protection against threats to the confidentiality, integrity, and availability of Kingdom resources and assets, Kingdom's information assurance policy currently addresses the basic security of the enterprise. It is essential that with new threats and vulnerabilities emerging every daythat the hospitals security posture is routinely assessed and updated to ensure that the hospitals enterprise network is not only secure but also in compliance with all financial, legal and health care compliance regulations. This is why an independent risk assessment was conducted for this case study will cover the entire hospital network including, but not limited to, remote clinics, wireless security, physical security and hospital security policy and compliance.

Kingdom enterprise hospital network security assessment was performed on several critical areas such as physical security (not only the network but also to sensitive areas such as the operating rooms, maternity wards and morgue), security management policies. The ultimate goal is to find the perfect balance between security and hospital operational functionality. Some areas where security weaknesses identified included the following;

- Shared network accounts: There are many network accounts that utilize a common username and password. This increases the security risk of unauthorized individuals gaining access to the network and does not provide for proper authentication.
- Weak authentication: the network user accounts only require a username and password with no password complexity required. Weak password policy is a key to network security because it is a common way attackers attack to gain access to a network.
- HIPAA Compliance: there is a lack of policy to ensure compliance to HIPAA Regulations.
- Acceptable use policy: no policy in place to communicate to the users what the acceptable behaviour is while utilising network resources, such as internet, email, social media and user responsibilities, increases various threats including viruses, malware and data loss.

### 6.1. KINGDOM VULNERABILITIES IDENTIFIED

During the security assessment, several identified threats and vulnerabilities pose could expose the hospital network to potentially dangerous security exposures. These vulnerabilities could open Kingdom hospital with damaging results, such as;

- Financial loss: a successful network attack allows an attacker the opportunity to manipulate data for financial gains.

- Loss of Reputation: Kingdom hospital success relies on it superior reputation with its customers and surrounding community, and a successful Denial of Service (DoS) attack or a breach leading to a compromise of sensitive information could lead to customers losing faith and confidence in Kingdom hospital.

- Legal consequences: A security breach leading to the compromise or loss of sensitive data opens up Kingdom hospital to legal issues.

### 6.2. Kingdom Enterprise Network Risks Identified

Mystical hospital security assessment has identified several as high-risk software and hardware risks to the enterprise network security. Some of these risks identified include;

- Software security: Software updates not implemented in a timely manner.
- Social media: Controls are not in place to ensure the users are fully aware of the potential dangers to Kingdom hospital from irresponsible or careless use of various social media sites.
- Malware/Virus: Controls are not in place to ensure that users are fully aware of how their irresponsible actions when utilizing internet and email can have dangerous effects on the network due to malware and/or virus infections.
- Mobile/Wireless devices: Controls are not in place to protect the wireless network again unauthorized mobile devices and/or individuals.
- Router/Switch: Configurations are default setting resulting in routers and switched being in an unsecure status.
- Firewall/IDS: Mystical does not have a firewall or IDS in place to protect the enterprise network from hostile attacks.

### 6.3. Kingdom Security Requirements

Kingdom's risk assessment has identified several security requirements that need to be address to ensure the protection of the patients, employees and various visitors at the hospital. Addressing these requirements will also protect the hospital assets, including financial and reputation in the surrounding area. Security requirements identified include;

- Natural Disasters: Being located in Mississippi the threat from hurricanes and tropical storms are a real and constant threat.
    - Mitigation: Perform quarterly review of Disaster Recovery/Business Continuity Plan with all required personnel on a quarterly basis.

- HIPAA Compliance: Being globally connected the threat from the compromise and/or loss of sensitive patient data is a real threat that requires special attention to ensure the protection of Personal Identification Information (PII), Protected Health Information (PHI) and Electronic Heath Records (EHR) Etc.
    - Mitigation: Have security measures in place to monitor email and ensure proper security measures are in place.

- Facility: It is essential to maintain constant electrical power and Air conditioning to maintain the critical hospital systems, such as servers and lifesaving medical equipment.
    - Mitigation: Ensure proper generators are in place and regularly tested in the event of power/HVAC loss.

- Insider threat: Intentional or unintentional, the insider threat posed a major risk to the network due to the potential for malware, virus or compromise of sensitive information.
    - Mitigation: Ensure all users receive proper training and measures are in place to monitor network for abnormalities.

### 6.4. Kingdom Security Training Policy

An important element of Kingdom security is the training and awareness program, because usually the first individual to see potential security issues is the end user. This IA training policy will ensure that all employees receive and maintain the proper level of IA training.Kingdom Hospital requires all individuals to attend an initial Information Training course and have a basic understanding of various concepts and principles to ensure the confidentiality, integrity and availability of Kingdom Hospital resources and assets, since they become our first line of defense. Some items that should be included in training;

- Recognizing unsafe email and/or attachments,
- Recognizing and avoiding potentially unsafe websites,
- Ensuring users understand the requirement to encrypt all emails containing sensitive information.
- Auditing and monitoring policy.

Table 2: Kingdom Hospital Risk Assessment Summary

| Kingdom Risk Assessment Summary | | | |
|---|---|---|---|
| Risk | Impact | Control/Mitigation Action | Early warning signs |
| Infrastructure | | | |
| default passwords on Routers/Switch | Depends | Ensure all devices have passwords changed from default. | Authorized systems denied logons, unauthorised devices accessing network. |

| Kingdom Risk Assessment Summary | | | |
|---|---|---|---|
| Risk | Impact | Control/Mitigation Action | Early warning signs |
| es lacking adequate Firewall | Depends | Upgrade to Firewall appropriate for enterprise. | Increased spam, malware or virus activity. |
| Natural Disaster | Low | Ensure disaster/contingency plan is developed and tested annually. | plan not regularly tested. |
| Network | | | |
| Shared Network Accounts | High | Enable network accounts to require role based smart card login or complex passwords if required. | accounts showing suspicious activity. |
| Weak Authentication | High | Enable all network user accounts to utilize smart card login. | User accounts login when user is not working or multiple logons at same time. |
| Unsecure mobile devices | High | Ensure mobile/wireless security is included in IA training, implement encryption on all mobile devices, Implement tracking software on all mobile devices, if possible. | Dataloss/compromise from loss of devices (mobile devices). |
| Compliance | | | |
| HIPAA/Privacy | High | Update policy to include HIPAA/Privacy requirements. Add additional training | Potential data breaches, sensitive information found accessible or sent through email. |
| System | | | |
| Regular Updates not completed | Medium | Ensure required updates are tested and completed in a timely manner. | Unauthorized access to systems/data. |
| Virus/Malware | Medium | Ensure users are educated about the dangers and ensure signatures are updated in a timely manner. | Applications not working properly, system freezes, unauthorized file changes, etc. |
| Training/Awareness | | | |
| Social Media | High | Ensure social media dangers are included in required IA training. | sensitive data being releases on social media sites. |
| Lack of Accountability | Depends | Requires users to acknowledge and sign Acceptable user agreement. | usersrepeat failure to comply with IA policy. |

## 7. CONCLUSION

Healthcare and healthcare IT has been advancing at an amazing rate in the last few years and while this had provided healthcare professional the ability to not only provide better care for their patients, it has also introduced new threats and vulnerabilities to information systems and sensitive patient information. Ensuring for the proper protection is an ongoing and challenging effort with its unique challenges, but with a properly developed and maintained IA program, it is possible. The ultimate goal is to understand the organizations business goals and objectives and find the proper balance between implementation of security controls, patient care and the business goals. These goals can be accomplished by prioritizing risks identified in the risk assessment and reduce those risks to an acceptable level. The U.S. Department of Health and Human Services has a website that gives the public information about medical ID theft and included is the Office of Inspector General's list of Most Wanted Health Care Fugitives (OIG)[14]. It is important that an organization never underestimate the users within the organization and like the Naval Intelligence motto; In God we trust, all others we monitor [15].

**Authors Bio:**

**Kimmarie Donahue** is a Senior Information Assurance Project Lead with KSH Solutions that provides Information Assurance, Information Security and Certification & Accreditation support services to various organizations within the federal government. She received her Master's Degree in Information Technology, Information Assurance and Security from Capella University. She currently holds the CISSP certification and NSA/CNSS (INFOSEC) Recognition. Kimmarie is a member of several professional organizations including ISC2, HIMSS, ISACA and ISSA.

**Syed (Shawon) M. Rahman** is an assistant professor in the Department of Computer Science and Engineering at the University of Hawaii-Hilo and an adjunct faculty of information Technology, information assurance and security at the Capella University. Dr. Rahman's research interests include software engineering education, data visualization, information assurance and security, web accessibility, and software testing and quality assurance. He has published more than 75 peer-reviewed papers. He is a member of many professional organizations including ACM, ASEE, ASQ, IEEE, and UPE.